\documentclass[
    ,final            
  ]
  {aipproc}

\usepackage[mathscr]{eucal}
 \usepackage{amsmath, amssymb}

 \def\msE{\mathscr{E}}

\newcommand{\be}{\begin{equation}}
\newcommand{\ee}{\end{equation}}
\newcommand{\bdm}{\begin{displaymath}}
\newcommand{\edm}{\end{displaymath}}
\newcommand{\vect}[1]{\mathbf{#1}}
\layoutstyle{6x9}

\begin{document}

\noindent {\it AIP Conference Proceedings, Vol.~1439, pp.~63--70 (2012)}\\
Proc. of WISAP\,2011 Conference ``Waves and Instabilities in Space and Astrophysical Plasmas'',
P.-L.\,Sulem \& M.\,Mond (eds.), Eilat, Israel,  June 19th  -  June 24th, 2011

\vspace{1cm}

\title{On the energy release in solar flares}


\author{L.A\,Pustil'nik}{
  address={Israel Space Weather and Cosmic Ray Center, Tel Aviv University, Israel Space Agency, \& Golan Research Institute, Israel, levpust@post.tau.ac.il}
}

\author{N.R.\,Ikhsanov}{
  address={Pulkovo Observatory, Pulkovskoe Shosse 65, Saint-Petersburg 196140, Russia}
}

\author{N.G.\,Beskrovnaya}{
  address={Pulkovo Observatory, Pulkovskoe Shosse 65, Saint-Petersburg 196140, Russia}
}

\begin{abstract}
High-resolution observations show the fine structure of the global equilibrium magnetic field configuration in solar atmosphere to be essentially different from that assumed in the traditional ``potential\,+\,force-free'' field scenarios. The interacting large-scale structures of fine field elements are separated by numerous non-force-free elements (tangential discontinuities) which are neglected in the traditional field picture. An incorporation of these elements into the model implies a dynamical rather than statical character of equilibrium of the field configuration. A transition of the system into flaring can be triggered by the ballooning mode of flute instability of prominences or/and coronal condensations. Tearing-mode and MHD instabilities as well as the effects of overheating of the turbulent current sheet prevent the field from stationary reconnection  as it is adopted in the traditional scenario. We speculate around the assumption that the energy release in active regions is governed by the same scenario as dynamical current percolation through a random  network of resistors in which the saltatory resistance is controlled by local current.
\end{abstract}

\maketitle


 \section{Introduction}

High resolution observations of solar atmosphere with space telescopes (e.g. SOHO, SDO, TRACE, HINODE, etc.) revealed that some parameters of the solar flares are inconsistent with the predictions of currently used theoretical models. This reopens the following questions:
 \begin{itemize}
\item What is the equilibrium configuration of the magnetic field and currents in the active regions before the flaring events? Is this a static or dynamical equilibrium and in the latter case, what is the topology of the phase trajectories attractor in the multi-dimension phase space of the system?
\item What could be the trigger of flares? Is the equilibrium system stable with respect to small-amplitude external perturbations?
\item How does the field and current structures evolve during the flare process and what is the mechanism of energy release? What is the post-flare configuration of the magnetic field?
 \end{itemize}
Possible answers to these questions are discussed in this paper.

  \section{Pre-flare field configuration}

Pre-flare equilibrium field configuration is usually treated in terms of a so called ``potential+force-free paradigm'' \cite{Priest-1984, Somov-2006}. The field configuration within this approach is considered as a superposition $\vect{H} = \vect{H}_0 + \vect{H}_{\rm ff}$, where $\vect{H}_0$ is the potential component ($\Delta \vect{H}_0 \equiv 0$) and $\vect{H}_{\rm ff}$ is the force-free component of the magnetic field, which satisfies the condition $\operatorname{rot}{\vect{H}_{\rm ff}} = \alpha \vect{H}_{\rm ff}$. Parameter $\alpha$ is accounting for the field helicity and in the general case can be inhomogeneous across the force tube of the magnetic field. The potential field can be unambiguously calculated if the boundary conditions (three components of the field at the photosphere) are specified. The task of determining the force-free component is more complicated. Only a limited number of analytical solutions have been derived for the most simple field configurations and under the assumption on an infinite half-space. Progress has been achieved in 2D numerical simulations. The boundary conditions in these calculations have been taken from observations and only a few simplifications about the structure and helicity of the magnetic field in the active region have been invoked (see e.g. \cite{Schrijver-etal-2008}, and references therein). A relatively good reproduction of the observed low-resolution maps of the magnetic field has been achieved for certain values of the free parameters of the model. Incorporation of more complicated boundary conditions (e.g. two or more dipoles or/and a dipole field in an external magnetic field) into the model has led, however, to singularities in which an unambiguous determination of the magnetic field is impossible. These singularities in 2D~calculation with non-static boundary conditions have been associated with regions in which the magnetic energy is accumulated and released during flares.



The above conclusion has been seriously challenged in studies of 3D~configurations in which the tangential component of the magnetic field is taken into account. The singular points in this case do not appear and the field configuration can be explained in terms of a magnetic field with shear, which is similar to the field configuration realized in TOKAMAK {\bf \cite{Kadomtsev-Shafranov-1983}}. A formation of a global static force-free magnetic configuration resulting from the interacting local force-free field structures is in this case impossible. Instead, the interaction between the local force-free field regions leads to a formation of tangential discontinuities \cite{Parker-1983} and re-construction of the global field configuration of a dynamically equilibrium state associated with the attractor in the phase space [$x_{\rm i}, \dot{x}_{\rm i}$]. This clarifies why the magnetic energy is accumulated in singular points in 2D simulations. This is a consequence of a basically incorrect assumption about the zero value of the magnetic field in the vicinity of these points and, correspondingly, the zero value of the magnetic tension (the energy of waves propagating in the vicinity of these points accumulates due to an infinitely large value of the refraction coefficient).

Analysis of high-resolution images (see e.g. the images of the sun taken by TRACE at http://soi.stanford.edu/results/SolPhys200/Schrijver/TRACEpodarchive.html) suggests that the solar atmosphere is threaded by a large number of narrow magnetic flux tubes connecting the chromosphere and the corona. The radius of the tube cross-section ranges from a few tens to a few hundred kilometers and remains almost constant along the tube from the bottom to the top. As the magnetic flux in the tubes is conserved, the field strength at the top of the tubes in the corona can be as high as hundreds (or even thousands) of Gauss. This indicates that the magnetic energy transferred from the photosphere to the corona is concentrated predominantly in the numerous narrow magnetic tubes which represent a global magnetic structure in dynamical equilibrium \cite{Parker-2001}. This significantly differs from the picture expected within the 'potential+force-free paradigm' (see e.g. \cite{Schrijver-Title-2011}, \cite{Ofman-etal-2011}).



Observations of the fine structure of the magnetic field in the solar photosphere
\cite{Stenflo-2004} show some evidence for the fractal structure of the magnetic field in active regions. This favors scenarios in which the fine structure is an intrinsic property of the magnetic field emerging from the convective region rather than a product of instabilities or/and turbulent motions in the solar atmosphere. Resulting pre-flare magnetic structure is an ensemble of numerous strongly interacting thin magnetic threads with avalanche-like percolation of free magnetic energy through this complex network \cite{Vlahos-Georgoulis-2004, Charbonneau-etal-2001,  Morales-Charbonneau-2010}.

  \section{Flare triggers and energy release processes}

 \subsection{Turbulent current sheet paradigm}

Solar flares are widely associated with anomalous dissipation of magnetic energy in the turbulent current sheet. Turbulization of the material in a region with strong current can be expected if the velocity of electrons exceeds the phase velocity of the plasma waves $u_{\rm curr} = j/ne > V_{\rm ph}$, where $j$ is the current density, and $n$ and $e$ are the number density and electric charge of the electrons. Otherwise, the waves will be suppressed by the Landau damping. The phase velocity of the plasma waves in the case of the ion-acoustic turbulence is $V_{\rm ph} = \left(T_{\rm e}/m_{\rm i}\right)^{1/2}$, where $T_{\rm e}$ is the electron temperature and $m_{\rm i}$ the ion mass. Hence, a formation of a turbulent current sheet in the solar atmosphere ($H \sim 100$\,G, $n \sim 10^8\,{\rm cm^{-3}}$, and $T \sim 10^6$\,K) could occur if the thickness of the current layer were 1--10\,km, i.e. a factor of $10^3 - 10^4$ smaller than the size of the active region itself. For the current sheet to remain active during the flare ($n \sim 10^{10}\,{\rm cm^{-3}}$, $T \sim 10^{7.5}$\,K), its thickness should not exceed 1--10\,cm. Under these conditions, however, the current sheet would be disintegrated by the tearing instability in a very short time. A question about the stability of the current sheet on the timescale of flaring events, therefore, remains open so far.

 \subsection{Trigger mechanisms}

The observed rapid transition of the system from its equilibrium state to a flare indicates that the magnetic configuration in which the energy is stored before the flare is metastable. In other words, the system before the flare is in a pre-catastrophe equilibrium in which a slight external perturbation, caused by the instability of a surrounding structure, may act as a trigger of flaring. Among the magnetic configurations which meets this criteria are the prominences and the coronal condensations. As shown in \cite{Pustilnik-1974, Pustilnik-1975}, these configurations are interchange unstable. In particular, a transition of these configurations into a flare can occur as the result of a flute instability. Observations of pre-flaring states have revealed that: (i)~Almost all flares are preceded by emerging of the magnetic flux which supplies the magnetic energy into the active region \cite{Ikhsanov-1968, Ikhsanov-1968a, Ikhsanov-etal-2004, Martres-etal-1963, Rust-1972}, (ii)~Some flares were observed to be preceded by oscillations of prominences in the vertical direction, expected at the first stage of the flute instability development \cite{Martin-Ramsey-1972, Pustilnik-1974}, (iii)~The strongest flares have been observed to be preceded by raising oscillations in coronal condensations \cite{Kobrin-etal-1976, Render-1983, Pustilnik-1978}, and (iv)~The flares develop as the phase wave propagates through the magnetic arches, starting from a local region of initial energy release. The magnetic energy in each of the arches is converted into plasma heating and particle acceleration up to relativistic energies. The energy of the particles propagating towards the solar surface is then converted into heating of chromospheric gas which fills the magnetic tubes and flows up to the corona, forming the coronal mass ejections. Thus, imbalance of unstable magnetic configurations by an additional magnetic flux emerging from the convective region may lead to a development of interchange instabilities of the prominences and coronal condensations and trigger the phase transition of the global magnetic structure.

  \section{Energy release in the current sheet}

It is presently well established that flares start with the formation of turbulent current sheet \cite{Kaplan-etal-1977}. However, the life-time of the current sheet is limited due to the following reasons.

{\bf Current sheet instabilities}. The characteristic time of tearing instability in the current sheets of thickness $d$, can be evaluated as $\tau_{\rm tg} \sim \tau_{\rm A}^{\rm s} \tau_{\rm d}^{\rm (1-s)}$, where $\tau_{\rm A} = d/V_{\rm A}$ is the Alfv\'en time and $V_{\rm A}$ the Alfv\'en velocity. $\tau_{\rm d} = 4 \pi \sigma d^2/c^2$ is the diffusion time and $\sigma$ is the plasma conductivity in the current sheet. Finally, the value of $s$ depends on the type of instability and lies in the range 0.25--0.5. The tearing instability of the current sheet is followed by pinch-type MHD instabilities (sausage and kink modes) on a time scale $\tau_{\rm pinch} \sim \tau_{\rm A}$, which leads to the disintegration of the current sheet.

{\bf Overheating of the current sheet}. A flare starts as turbulization of the plasma in the current sheet occurs. The onset condition for turbulization reads $j > j_{\rm cr}$, where $j_{\rm cr}= neu_{\rm cr}$ is the critical current density and $u_{\rm cr}$ the critical velocity, which is equal to $V_{\rm T_{\rm e}} = \left(k T/m_{\rm e}\right)^{1/2}$ for $T_{\rm e} = T_{\rm i} = T$, or $c_{\rm s_{\rm i}} = \left(k T_{\rm e}/m_{\rm i}\right)^{1/2}$ for $T_{\rm e} \gg T_{\rm i}$ \cite{Pustilnik-1975}. Here $n$ is the number density and $T_{\rm e}$ and $T_{\rm i}$ are the electron and ion temperatures in the current sheet. As the turbulization starts, the temperature of plasma in the current sheet increases on a time scale of $\sim 1/\omega_{\rm 0i}$, which under the conditions of interest is about a few microseconds (here $\omega_{\rm 0i}$ is the ion plasma frequency). If the cooling of the plasma in the current sheet is ineffective, the gas temperature reaches a critical value at which the onset condition for plasma turbulization is no longer satisfied and plasma waves are suppressed by Landau damping. It, therefore, appears that a stationary reconnection process can operate in the turbulent current sheet only if cooling dominates heating in the reconnection region \cite{Pustilnik-1975}. This is unlikely to be satisfied if the scale of the current sheet along the field lines is comparable with the size of the active region. This problem could, however, be avoided if the current sheet were disintegrated into numerous fragments in which the energy dissipation occurs. The field configuration in this case would, however, be essentially different from that used in the traditional model, and the question about the reconnection mechanism in this situation remains open.

It, therefore, turns out that both theoretical analysis and observations suggest that flares in solar atmosphere operate in a form of transition waves, which propagate through the system in a state of catastrophic equilibrium. The wave triggers the energy release process in local regions, and also provides a feed-back to the whole system. Next section shows that such a behavior is rather common and drives many puzzling processes observed on the Earth.

 \section{Flares as current percolation through a random network of resistors}

\begin{figure}
  \includegraphics[height=.65\textheight,angle=90,]{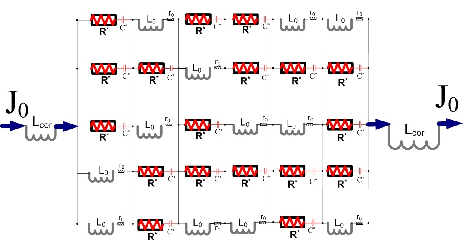}
  \caption{Percolation of the electric current, $J_0$, through a network of resistors. $L_0$ and $r_0$ are the inductivity and resistance of ``normal'' high conductive elements. $R^*$ and $C^*$ are the abnormal resistance and electrostatic double layers of turbulent low conductive elements}
   \label{f3}
\end{figure}

Plasma instabilities and overheating of the turbulent current sheet lead to a disruption of the magnetic structure in active regions into a random network of ``normal'' and ``turbulent'' domains. A qualitative model of these domains can be constructed in terms of resistor elements (see Fig.\,\ref{f3}). Under the condition of global current conservation (which is a natural consequence of high inductivity of the corona where the current sheets are located) the current is forced to percolate through this  network of resistors \cite{Pustilnik-1998}. Since conductivity of the resistors depends on the current density, $j_{\rm ik}$, a random  redistribution of the local currents in this network occurs. This redistribution is accompanied by turbulization of new elements as the current density in a local region increases over a critical value, $j_{\rm ik} > j_{\rm cr}$. The system, therefore, switches into a dynamically equilibrium phase being governed by a non-linear feedback from one element to another through  variations of the resistivity of the individual elements and currents redistribution.

This system can be described in terms of the percolation in a random network of resistors in which the resistance of each element, $R_{\rm ik}$, depends on the local current density $j_{\rm ik}$ as:
 \be
 \left\{
  \begin{array}{lcl}
R_{\rm ik} = R_0 & {\rm for} & j_{\rm ik} \leq j_{\rm cr} \\
R_{\rm ik} = R_{\rm t} \sim 10^6\,R_0 & {\rm for} &  j_{\rm ik} > j_{\rm cr}, \\
   \end{array}
\right.
 \ee
where $R_0$ is a very low ``normal'' resistance of ``good'' elements and $R_{\rm t}$ is a very high ``abnormal'' resistance of ``bad'' elements in the network. Distribution of currents and voltage on the elements of the network is controlled by Kirchhoff laws. In particular, $J_{\rm k} = \sum j_{\rm ik} = J_0$ due to the conservation of the total current, $J_0$, at any $k$-section of the network while the voltage along any $k$-circuit with the same initial and final points in the network is $\Delta V_{\rm i} = \sum j_{\rm ik} R_{\rm k}$.

Note that percolation through random network of conductive elements (or fractal clusters of elements) has a wide range of applications \cite{Stauffer-Aharony-1999, Feder-1988} from creation of nanomaterials with given conductive and optical properties to the study of spreading of forest fires, infection deceases and magnetospheric storms \cite{Zelenyi-Milovanov-2004}. This process was studied in direct experiments (in superconductive ceramic samples \cite{Vedernikov-etal-1994} and conductive graphite paper with random holes \cite{Levinshtein-etal-1976}) in numerical simulations \cite{Kirkpatrick-1973} and analytically \cite{Render-1983}). However, a main tool to study percolating systems is computer simulation, taking into account the character of interconnection between the elements, dimension of the lattice and fractal dimension of clusters formed by conductive elements.



The percolating systems have the following common properties: (i)~Phase transition of the system at a certain value of critical parameter usually referred to as a percolation threshold manifests itself in drastic changes of the system properties (see e.g. \cite{Epstein-1987}), (ii)~Self-clustering of the network's elements into fractal-like patterns with a power-law spectrum of basic parameters, and (iii)~Non-linear behavior of the system due to feedback from individual elements.

One of the key properties of the network is the threshold-like character of the global resistance $R_{\rm net}$, which depends on the density of ``bad'' resistive elements, $p_{\rm i}= p_{\rm i}(J_0)$, and on the total current in the turbulent current sheet, as $R_{\rm net} \propto (J_0 - J_{\rm cr})^{-\alpha}$. The flaring starts as $R_{\rm net} \rightarrow \infty$. This is expected if the density of the ``bad'' elements increases or/and $J_0 \rightarrow J_{\rm cr}$, at which the phase transition of the system occurs.

A particular property of the percolating random network is that the number of current clusters, $N(x)$, depends on the network parameters (such as the size, volume, length, etc.) as $N(x) \propto x^{-n}$. This indicates that the amplitude-frequency spectrum of solar flares within this approach must be a power-law, $N(W) \propto W^{-m}$, and the same spectrum is expected for flares in red dwarfs of UV~Cety--type.

The same is valid for particles accelerated in clusters of the random network of resistors. The electric field generated during flaring events in the clusters can be evaluated as $E_* = \dfrac{j_{\rm cr}}{\sigma_*} = (10^{-2} - 10^{-3}) \dfrac{n e c_{\rm s_{\rm i}}}{\omega_{\rm 0i}}$, where
$\sigma_* = \dfrac{\omega_{\rm 0e}^2}{4 \pi \nu_{\rm eff}} \sim (10^2 - 10^3) \omega_{\rm 0i}$
is the conductivity in the flaring region under condition $j > j_{\rm cr} = n e c_{\rm s_{\rm i}}$. Here $c_{\rm s_{\rm i}}$ is the ion sound speed and $\omega_{\rm 0e}$ is the electron plasma frequency. The energy of particles accelerated by the electric field, $\msE \sim e E_* l_{\rm z}$, can be as high as
 \be
 \msE \sim e E_* L = (1 - 10)\,{\rm GeV} \times  n_8^{1/2} T_7^{1/2} \left(\frac{l_{\rm z}}{10^9\,{\rm cm}}\right),
 \ee
where $n_8 = n/10^8\,{\rm cm^{-3}}$ and $T_7 = T/10^7$\,K. Since the number of clusters of the length $l_{\rm z}$ in the network region is $N(l_{\rm z}) \propto l_{\rm z}^{-k}$ the expected energy spectrum of the accelerated particles is $N(\msE = e E_* l_{\rm z}) \propto \msE^{-k}$, where the value of $k$ depends on the basic properties of the network. This indicates that the percolating network scenario can provide us with a natural explanation of the observed power-law spectrum of solar cosmic rays.

 \section{Conclusions}

The pre-flaring fine structure of the magnetic field in active regions observed with space missions differs from the static ``potential+force-free'' field configuration assumed in the traditional scenarios of flares. It appears that the field configuration is in a dynamical rather than static equilibrium and resembles the configuration of the field with shear observed in TOKAMAKs. Plasma instabilities and overheating of the turbulent current sheet prevent the energy release process from  operating steadily and disintegrate the current sheet into numerous turbulent and ``normal'' domains. The dynamical equilibrium of this system, which is controlled by the level of global current can be explained in terms of the current percolation through a random network of resistors. This makes it possible to explain the power-law statistical properties of flares and microflares and particle acceleration with the power-law energy spectrum.

\begin{theacknowledgments}
LAP and NRI thank the organizing committee for kind hospitality and French Embassy in Israel (Office of Science and Technology) and Ben-Gurion University of the Negev for support in attending the Workshop. The research has been partly supported by the Program of RAS Presidium N\,19 and NSH-3645.2010.2, and by the grant ``Infrastructure'' of Israel Ministry of Science and COST~ES~0803.
\end{theacknowledgments}

\end{document}